\documentstyle[12pt]{article}

\newcounter{saveeqn}
\newcommand{\alpheqn}{\setcounter{saveeqn}{\value{equation}}%
\stepcounter{saveeqn}\setcounter{equation}{0}%
\renewcommand{\theequation}{\mbox{\arabic{saveeqn}-\alph{equation}}}}
\newcommand{\reseteqn}{\setcounter{equation}{\value{saveeqn}}%
\renewcommand{\theequation}{\arabic{equation}}}

\begin{document}
\begin{titlepage}

\hskip 12cm 
\vskip 0.6cm
\centerline{\bf PARAMETRISING THE PROTON STRUCTURE FUNCTION}
\vskip 1.0cm
\centerline{   L. L. Jenkovszky$^{a\dagger}$, A. Lengyel$^{b\ddagger}$, F. 
Paccanoni$^{c\ast}$}
\vskip .5cm
\centerline{$^{a}$ \sl  Bogoliubov Institute for Theoretical Physics,}
\centerline{ \sl Academy of Sciences of the Ukrain}
\centerline{\sl 252143 Kiev, Ukrain}
\vskip .5cm
\centerline{$^{b}$ \sl Institute of Electron Physics}
\centerline{\sl Uzhgorod 294016, Universitetska 21, Ukraine}
\vskip .5cm\centerline{$^{c}$ \sl  Dipartimento di Fisica, Universit\`a di Padova,}
\centerline{\sl Istituto Nazionale di Fisica Nucleare, Sezione di Padova}
\centerline{\sl via F. Marzolo 8, I-35131 Padova, Italy}
\vskip 1cm
\begin{abstract}
We show that simple parametrisations at small $x$ of the proton
structure function work so well in limited regions of the $(x, Q^2)$
plane because they are approximately "self-consistent" solutions
of the QCD evolution equation. For a class of them, we predict
their $Q^2$ dependence and compare the result with experimental
data.

\end{abstract}
\vskip .5cm
\hrule
\vskip.3cm
\noindent

\vfill
$\begin{array}{ll}
^{\dagger}\mbox{{\it email address:}} &
   \mbox{JENK@GLUK.APC.ORG}
\end{array}
$

$ \begin{array}{ll}
^{\ddagger}\mbox{{\it email address:}} &
 \mbox{SASHA@LEN.UZHGOROD.UA}
\end{array}
$

$ \begin{array}{ll}
^{\ast}\mbox{{\it email address:}} &
   \mbox{PACCANONI~@PADOVA.INFN.IT}
\end{array}
$
\vfill
\end{titlepage}
\eject
\textheight 210mm
\topmargin 2mm
\baselineskip=24pt
{\bf 1. Introduction}

Since long time non-Regge behaviour of the proton structure function
$F_2(x,Q^2)$ in deep inelastic scattering~\cite{ADR} has been
discussed on the basis of the scale break predicted from QCD.
Recent experimental data~\cite{H1,ZEUS} provide further evidence
for perturbative QCD evolution equations~\cite{GL,AP} and
confirm~\cite{H1} the scaling behaviour of the singlet proton
structure function at small $x$ and large $Q^2$ predicted in
ref.~\cite{BF}. This double asymptotic scaling is derived from
the exact solution of the dynamical equations~\cite{AP} in the
region where nonsinglet contributions to $F_2(x,Q^2)$ can be
neglected and the lowest moments of the splitting functions
dominate.
\vskip.3cm
Empirical expressions for $F_2(x,Q^2)$, where the functional form 
of the perturbative double scaling is respected, will
reproduce well experimental data. An example is the parametrisation
of ref.~\cite{RW}, where an interesting similarity in the
energy dependence between the average charged multiplicity in 
$e^+\,e^-$ collisions and the proton structure function has been
exploited.
\vskip.3cm
Attempts to implement the unitarity condition in deep inelastic
scattering~\cite{BAL,FP,BH} lead to a quite different behaviour
for the proton structure function. As suggested in~\cite{SF}
double scaling admits local approximations due to the slow
variation with $x$ and $Q^2$ of the relevant variables in the
asymptotic region. We agree with this interpretation but other
reasons can be found for the agreement of these simple
parametrisations with HERA data.
\vskip.3cm
If we consider two simple expressions for $F_2(x,Q^2)$:
\begin{equation}
F_2(x,Q^2)=a(Q^2)+b(Q^2) \left(\frac{x_0}{x}\right)^{c(Q^2)}
\label{z1}
\end{equation}
and
\begin{equation}
F_2(x,Q^2)=\sum_{i=0}^2 a_i(Q^2) \left(\ln\frac{x_0}{x}\right)^i
\label{z2}
\end{equation}
and fit them at low $x$, for each $Q^2$ bin, to the experimental
data~\cite{H1,ZEUS}, we find that ansatz (\ref{z2}) has a lower
total $\chi^2_{d.o.f.}$, while the exponent $c(Q^2)$ in (\ref{z1}) 
agrees with the finding of~\cite{H1}. The good description of
data obtained with the parametrisation (\ref{z2}) must find
an explanation in terms of the evolution equation~\cite{AP}.
The purpose of this paper is to find a reason for this agreement.
\vskip.3cm
In Section {\bf 2} we will discuss new approximate solutions of the
standard evolution equations~\cite{AP} in a form similar to the
ones found in~\cite{BAL,FP,BH}. By limiting ourselves to the
double logarithmic approximation, an explicit example will be given.
In Section {\bf 3} we will show that there is agreement between
these solutions and an explicit fit to the data. Last Section is
devoted to a discussion of the results.

\vskip0.5cm

{\bf 2. A theoretical explanation.}

Let us consider the singlet evolution equation~\cite{AP} at 
leading order, when only the largest singularity in moment
space contribute to the splitting functions. Neglecting non 
singlet contributions and defining the new variables
\begin{equation}
\xi=\ln\left(\frac{x_0}{x}\right), \mbox{and    }
\zeta=\ln\left(\frac{\ln(Q^2/\Lambda^2)}{\ln(Q_0^2/\Lambda^2)}\right)
\label{z3}
\end{equation}
a simplified form for the problem of determining the proton 
structure function can be derived~\cite{BF} in the double
logarithmic approximation
\alpheqn
\begin{equation}
F_2^p=\frac{5}{18}Q(\xi,\zeta)
\label{z4a}
\end{equation}
\begin{equation}
Q(\xi,\zeta)=Q(\xi,0)+\frac{f\gamma^2}{9}\int_0^{\zeta}\,d\zeta '
G(\xi,\zeta ')
\label{z4b}
\end{equation}
\begin{equation}
\frac{\partial^2 G(\xi,\zeta)}{\partial\xi \partial\zeta}+
\delta \frac{\partial G(\xi,\zeta)}{\partial\xi}-\gamma^2 G(\xi,
\zeta)=0
\label{z4c}
\end{equation}
\reseteqn
\noindent
where $Q(\xi,\zeta)$ and $G(\xi,\zeta)$ are the singlet quark
distribution and the gluon distribution multiplied by $x$,
respectively. In the equations above, $f$ is the number of flavours
and 
\begin{equation}
\gamma=\sqrt{\frac{12}{\beta_0}},\mbox{   } \delta=\left( 11+
\frac{2 f}{27}\right)/\beta_0
\label{z5}
\end{equation}
with $\beta_0=11-2f/3$.
\vskip.3cm
As well known~\cite{FT} the Riemann function for the Goursat
problem in eq.(\ref{z4c}) is
\begin{equation}
U(\xi,\zeta;\xi ',\zeta ')=e^{\delta (\zeta-\zeta ')}
I_0[2\gamma\sqrt{(\xi'-\xi)(\zeta '-\zeta)}]
\label{z6}
\end{equation}
where $I_0(z)$ is the modified Bessel function of order zero.
The solution of (\ref{z4c}) can be written
in the form
\begin{equation}
G(P)=(UG)_A+\int_A^{P_1}\,d\zeta\,U\left(\delta G+
\frac{\partial G}{\partial\zeta}\right)+\int_A^{P_2}\,d\xi
U\frac{\partial G}{\partial \xi}
\label{z7}
\end{equation}
where $P=(\xi ',\zeta '), P_1=(\xi_0,\zeta '),
P_2=(\xi ', \zeta_0)$ and
$A$ is the point $(\xi_0, \zeta_0)$ of the $(\xi', \zeta')$ plane.
\vskip.3cm
The asymptotic form of $I_0(z)$, for large argument, leads
from (\ref{z7}) to the double asymptotic scaling~\cite{BF},
for soft boundary conditions along $AP_1$ and $AP_2$.
In the asymptotic region, starting scales $x_0$ and $Q_0^2$
are not important and a lower bound on $\sqrt{\xi\zeta}$ has
been imposed in~\cite{BF} in order to get scaling in this variable.
\vskip.3cm
Near the origin of the $(\xi,\zeta)$ plane, eqs.(\ref{z4a},
\ref{z4b}) give a poor approximation of the reality, if $x_0$
is not small enough, because the non singlet contribution 
becomes sizable. We can try to avoid in part the
valence quark region by imposing on the experimental data the
cuts: $x\leq 0.05$ and $y>0.02$ as in ref.~\cite{RW}.
Once $x_0$ has been fixed, $x_0=0.05$, with the choice 
$Q_0^2=6.5\,GeV^2$ H1~\cite{H1} and ZEUS~\cite{ZEUS} data
cover a small strip of the $(\xi,\zeta)$ plane. The
maximum value of $y=Q^2/(xs)$ is 0.56 for H1 and 0.9 for
ZEUS. $s=90200\,GeV^2$ is the square of the center of mass
energy of the lepton-proton collision. With $\Lambda=
200\,MeV$ the aforesaid bounds can be translated in the
variables $\xi,\zeta$
\begin{equation}
5.1 e^{\zeta}+\xi \approx \ln\left(\frac{x_0ys}{\Lambda^2}
\right)
\label{za}
\end{equation}
and it turns out that, for three (four) flavours, the argument
of the Bessel function in (\ref{z6}) reaches a maximum value
of 2.8 (3.0) for ZEUS and 2.6 (2.8) for H1.
\vskip.3cm
In all the above region the series expansion for $I_0(z)$, 
truncated to the first three terms, provides an approximation
better than the standard asymptotic expansion for all $z$,
up to $z\approx 2.8$. However, already at $z=4$, the relative
errors become 20 and 3.6 percent, respectively. This can 
explain both the limitations of this truncation and
the interest for a local parametrisation whose success is 
doomed to an abrupt failure. Outside the region outlined above,
deviations from QCD perturbative evolution are in fact under
control in the simple expression for $F_2(x,Q^2)$ we will
obtain. The onset of new dynamical effects, like damping
by screening corrections~\cite{GLR}, can be easily detected
and, finally, a reasonable and simple input for evolution is
suggested, just at the border of the existing published
data.
\vskip.3cm
We return now to eq.(\ref{z7}). Boundary conditions can be given
along the lines $\zeta_0=0$ and $\xi_0\approx 2.75$ (3.0) for
H1 (ZEUS). This choice satisfies the cut (\ref{za}) and is
equivalent to a shift in the starting point $x_0$ as far as
the evolution is concerned. A more refined solution, where
part of the boundary is given by eq.(\ref{za}) with $y=0.02$,
would meet with difficulties in the choice of the initial
distributions. Since the series for $I_0(z)$ will be truncated, 
$I_0(2\sqrt{z})\approx 1+z+z^2/4$, the evaluation of $G(\xi,\zeta)$, 
for given boundary conditions, is immediate. If the starting 
conditions are
\begin{equation}
G(\xi,0)=\alpha_1+\lambda (\xi-\xi_0)
\label{z8}
\end{equation}
and
\begin{equation}
G(\xi_0,\zeta)=\alpha_1+\alpha_2 \zeta
\label{z9}
\end{equation}
with a n-terms approximation for $I_0(z)$, we end up with a
polynomial in $\xi$, of degree (n+1), for $G(\xi,\zeta)$.
We argued that a three terms expansion for $I_0(z)$ is needed
to reproduce correctly the $Q^2$ evolution, then $G(\xi,\zeta)$,
and hence $F_2$, should be a polynomial of fourth degree in the
variable $\xi$ with coefficients depending on $\zeta$.
\vskip.3cm
In order to reduce the number of free parameters
we can perform the last integral in eq.(\ref{z7}) obtaining
\begin{displaymath}
\lambda\frac{e^{-\delta\zeta'}}{\gamma}\sqrt{\frac{\xi'-\xi_0}{\zeta'}}
I_1[2\gamma\sqrt{\zeta'(\xi'-\xi_0)}]
\end{displaymath}
and then approximate the integral in eq.(\ref{z4b})
in the form
\begin{displaymath}
\int_0^{\zeta'}\,\frac{e^{-\delta\zeta}}{\sqrt{\zeta}}
I_1[2\gamma\sqrt{(\xi'-\xi_0)\zeta}]\,d\zeta \approx
\end{displaymath}
\begin{equation}
\gamma\sqrt{\xi'-\xi_0}\left(\frac{1}{\delta} (1-e^{-\delta \zeta'})
+\frac{\gamma^2(\xi'-\xi_0)}{2}(\frac{1}{\delta}-e^{-\delta\zeta'}
(\zeta'+\frac{1}{\delta}))\right)
\label{z10}
\end{equation}
Errors in eq.(\ref{z10}) have been checked numerically and are
compatible with the approximation adopted. At high $Q^2$, the result
will lose in accuracy but, with this expedient, we can utilize in
the fit the $Q^2$ bins where few data appear.
\vskip.3cm
If we rewrite eq.(\ref{z2}) in the form
\begin{equation}
F_2=\frac{5}{18}\sum_{i=0}^2\,c_i(\zeta)(\xi-\xi_0)^i
\label{z12}
\end{equation}
we get the relations $(c_i=c_i(0))$
\alpheqn
\begin{equation}
c_0(\zeta)=c_0+\frac{\gamma^2 f}{9}(\alpha_1\zeta+\alpha_2
\frac{\zeta^2}{2})
\label{z13a}
\end{equation}
\begin{equation}
c_1(\zeta)=c_1+\frac{\gamma^4 f}{9\delta^2}\left[ \left(-\alpha_1+
\frac{\alpha_2}{\delta}+\frac{\delta\lambda}{\gamma^2}\right)
\left(1-e^{-\delta\zeta}\right)+(\delta\alpha_1-\alpha_2)\zeta +
\frac{\delta\alpha_2\zeta^2}{2}\right] 
\label{z13b}
\end{equation}
\begin{displaymath}
c_2(\zeta)=c_2+\frac{\gamma^6 f}{18\delta^3}\left[\left(
-2\alpha_1+\frac{3\alpha_2}{\delta}+\frac{\delta^2\lambda}{\gamma^2}
\right) \left(1-e^{-\delta\zeta}\right) \right. +
\end{displaymath}
\begin{equation}
\left. +(\delta\alpha_1-2\alpha_2)\zeta+\left(\delta\alpha_1-
\alpha_2-\frac{\delta^3\lambda}{\gamma^2}\right)\zeta e^{-\delta
\zeta} +\frac{\delta\alpha_2\zeta^2}{2} \right]
\label{z13c}
\end{equation}
\reseteqn
\vskip.3cm
Boundary conditions are imposed by giving $c_0(\zeta)$ (hence
$c_0$, $\alpha_1$ and $\alpha_2$ are known) and the constants 
$c_i$ ($i=1,2$). Finally the constraint
\begin{displaymath}
\left. \frac{\partial c_1(\zeta)}{\partial \zeta}\right|_{\zeta=0}
=\frac{\gamma^2 f}{9} \lambda
\end{displaymath}
determines $\lambda$ if the derivative of $c_1(\zeta)$, for
$\zeta=0$, is known from experiment.
The application of the above condition is difficult since
experimental data have large errors. However gluon density
must satisfy other approximate constraints ~\cite{KP,EKL,GLUE} and,
since it is universal,
its behaviour can be inferred also from vector meson~\cite{MGR} and
diffractive jet~\cite{NZ} production.
\vskip.3cm
We notice that the initial conditions (\ref{z8}) and (\ref{z9})
must be considered as an example since (\ref{z8}) should be
multiplied by a factor $(1-x)^{\alpha}$, that is $(1-x_0
e^{-\xi})^{\alpha}$, and (\ref{z9}) is perhaps too naive. As a
consequence of the neglect of the term $(1-x)^{\alpha}$,
predictions for $F_2(x,Q^2)$ will be above the experimental
data for the points with large $x$. The effect of this approximation
will be in fact more important than the neglect of the
valence quarks contribution in the selected region.
\vskip 0.5cm

{\bf 3. Comparison with a fit to the data.}
 
We argued that eqs.(\ref{z12}) and (\ref{z13a}-\ref{z13c}) will
reproduce the experimental proton structure function in the
region of the $(x,Q^2)$ plane specified above. To substantiate
this belief we performed a fit of the experimental data, for
each $Q^2$ bin separately, to the functional form (\ref{z12}).
\vskip.3cm
To facilitate the comparison with the theoretical predictions,
$\xi_0$ in (\ref{z12}) has been chosen as $\xi_0=\ln(0.05/
0.0032)$ for H1 data~\cite{H1} that will be considered first.
The total $\chi^2_{d.o.f.}$ for all the 16 values of $Q^2$,
from $6.5\,GeV^2$ to $500\,GeV^2$, is 3.9 using the full
errors, with a mean value of 0.24 for each $Q^2$ bin fitted.
Parametrisation (\ref{z1}) would give instead a total
$\chi^2_{d.o.f.}$ of 4.7.
This finding justifies our claim in the Introduction that
parametrisation (\ref{z2}) is preferable to (\ref{z1}).
\vskip.3cm
The result from the fit is displayed in fig.1, where the values
of the coefficients in eq.(\ref{z12})
\begin{equation}
  A_i(\zeta)=\frac{5}{18} c_i(\zeta) \mbox{     }
(i=0,1,2)
\label{z14}
\end{equation}
have been plotted with closed circles as function of $Q^2$.
The continuous curves in fig.1 are obtained from eq.
(\ref{z13a}-\ref{z13c}) with the following boundary
conditions
\begin{displaymath}
  c_0=2.24,\;c_1=0.46,\;c_2=0.063
\end{displaymath}
\begin{equation}
\alpha_1=5.17,\;\alpha_2=9.83,\;\lambda=2.53
\label{z15}
\end{equation}
and $f=4$. While the parameters $\alpha_1$ and $\alpha_2$
have been fixed from a fit to $F_2(x=0.0032,Q^2)$, eq.(
\ref{z13a}), a guess for $\lambda$ has been obtained from
the experimental gluon momentum densities~\cite{H1,GLUE}.
\vskip.3cm
In fig.2 the continuous curves show the gluon momentum
density as function of $x$, obtained with the parameters in 
(\ref{z15}), at $Q^2=6.5\,GeV^2$ and evolved at $Q^2=20\,GeV^2$,
without any approximation on the last integral in eq.(\ref{z7}).
Comparison with the proposal in~\cite{H1,GLUE} shows that 
the slope $\lambda$ is almost correct, while the value of
$\alpha_1$ is below what one would expect by a 20 percent.
This discrepancy can be ascribed to the difficulty of
fitting the proton structure function, at fixed $x$, as
function of $Q^2$. Both the simplified input and the large
errors in the data contribute to the difference that can be
quantified in a leading-order calculation. The dashed curve
in fig.2 represents the result for $G(x,Q^2)$ obtained from
eq.(\ref{z12}) with the leading order method of ref.
~\cite{KP}. Next-to-leading order calculation would lower
this prediction, but the value for $\alpha_1$, suggested
by the fit, seems really too small.
\vskip.3cm
The same fit has been repeated for the ZEUS data~\cite{ZEUS}.
The total $\chi^2_{d.o.f.}$ increases, with a mean value of 
0.8 for each $Q^2$ bin fitted. $A_0(\zeta)$ and $A_1(\zeta)$
appear as in fig.1, but the fit for $A_2(\zeta)$ deteriorates 
somewhat. While the spread of the fitted points increases,
the curves defined in eqs.(\ref{z13a}-\ref{z13c}) represent
always well the general behaviour. The different regions of
the $(\xi,\zeta)$ plane covered by the two sets of data
account only partly for the difference in the two fits.
However, other parametrisations presented us with the same
problem.
\vskip.3cm
Going back to the H1 data, examples of the proton structure
function $F_2(x,Q^2)$, obtained with this method, are given
in fig.3 with continuous curves, at different $Q^2$ values,
as function of the variable $\sqrt{s_{eq}}$ introduced in
ref.~\cite{RW},
\begin{displaymath}
 \sqrt{s_{eq}}=\frac{2Q_1}{\sqrt{x}}.
\end{displaymath}
The choice of this variable, where $Q_1=270\,MeV$, allows for
a comparison of the energy dependence of the mean charged
multiplicity with the $x$ behaviour of $F_2$.
\vskip.3cm
Quadratic fits in $\ln s$ to the multiplicity data for
$p-p$ scattering are rather old~\cite{ALB,THO} and their
derivation in the framework of a Regge model has been
suggested~\cite{KMS}. Till 1992 the phenomenological fit
of ref.~\cite{THO} has been shown in the Review of
Particle Properties~\cite{HIK} and, from then, an analogous
version adapted to $e^+-e^-$ annihilation became 
available~\cite{ACT}.
\vskip.3cm
In fig.3 we have superimposed to the data also the predictions
for the mean charged multiplicity of ref.~\cite{THO}
(dashed line) and of ref.~\cite{ACT} (dotted line) normalized
to the $F_2$ data at each $Q^2$ value. The idea of
similarity between the two observables, advanced in ref.
\cite{RW}, seems confirmed in this more qualitative approach.
Remarkably, this similarity could be process independent at
high energy. Notice that~\cite{ACT} the empirical logarithmic 
solution has a better $\chi^2_{d.o.f.}$ than the leading-log
approximation proposed in~\cite{FPP} and provides a fit to
the mean charged multiplicity comparable to the modified
leading-log approximation~\cite{KO} in the energy range
under consideration.
\vskip 0.5cm

{\bf 4. Conclusions}

Motivations for the success of simple parametrisations for 
the proton structure function have been found in the 
structure of the QCD evolution equations. In this paper
we have shown that a region of the $(x,Q^2)$ plane exists
where both asymptotic and truncated series expansions
give the answer with a comparable accuracy. In the double
logarithmic approximation, we are considering, this region
encloses almost all HERA data while, at smaller $x$,
the asymptotic solution would certainly be the only
appropriate.
\vskip.3cm
Except for an exact next-to-leading order computer
calculation, the logarithmic fit appears to be at the level of 
other simple parametrisations, at least in the explored domain of
the relevant variables. This phenomenological finding
has a correspondence in the theoretical picture of deep
inelastic scattering, where $F_2(x,Q^2)$ is obtained from
the discontinuity of the Pomeron exchanged between the
proton and a quark loop coupled to the photon.
Representing the Pomeron as a gluonic ladder, the Born
term, correponding to the exchange of two gluons, would
produce roughly an $x$ independent asymptotic 
structure function. Every rung added to the ladder brings
about a term increasing like $\ln(1/x)$ when $x$ tends
to zero~\cite{CW}. Hence, a two-rung gluonic ladder is 
suggested from our logarithmic fit.
\vskip.3cm
Two important points must however be noticed. First, at the
border of the $(\xi,\zeta)$ region studied, this approach
shows his limitations and the fit deteriorates somewhat.
The second point regards the possibility to extend the 
range of validity of this method by keeping more terms in
the expansion for the Bessel function. The answer is negative.
Mathematical tables~\cite{AS} show that, in the double
logarithmic approximation, double scaling becomes
asymptotically ineluctable.

\newpage

\centerline{\bf Figure Captions}
\vskip .3 cm
\begin{description}

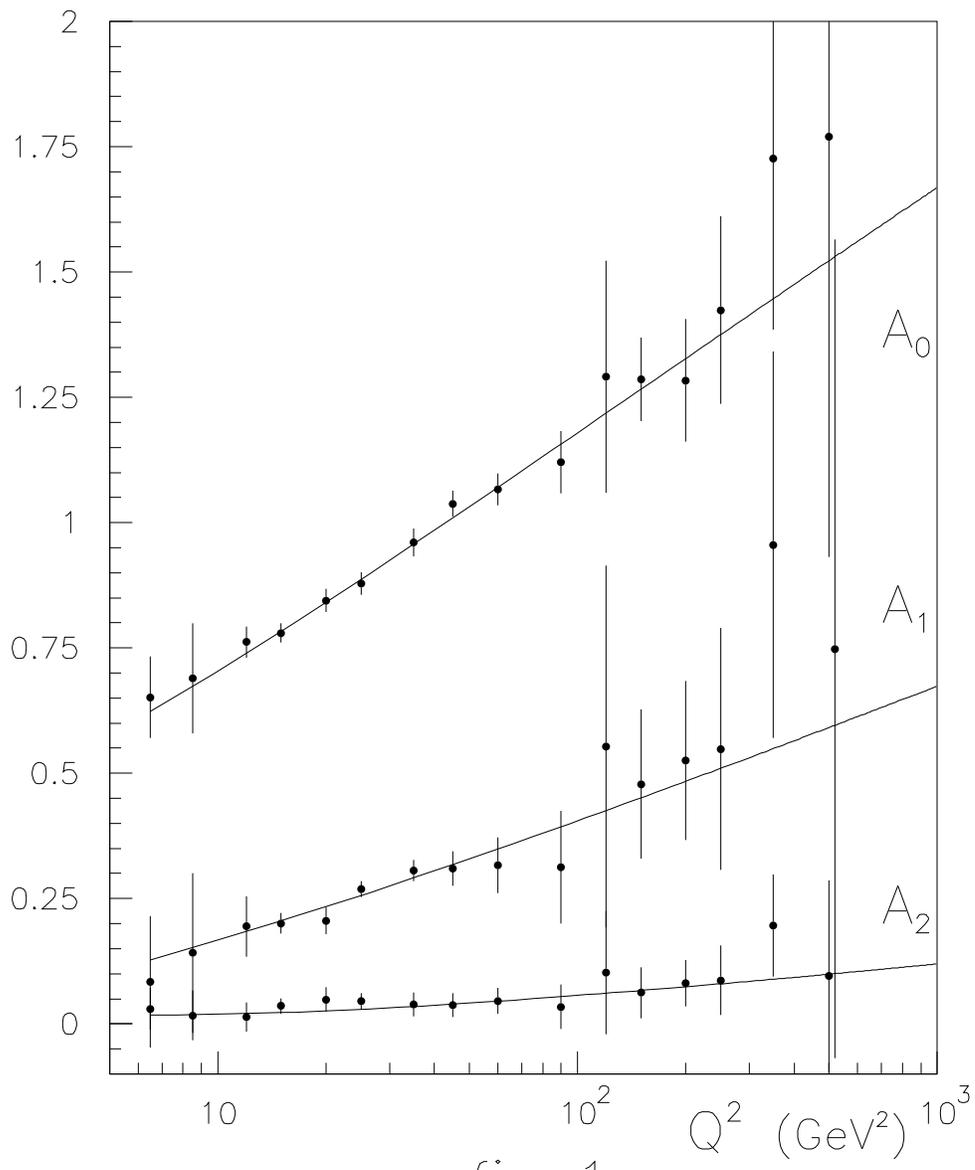
\item{fig. 1:} Coefficients of the logarithmic fit, plotted as
function of $Q^2$for the data in~\cite{H1}, the continuous line
shows the prediction of the model (see text). The point at 
$500\,GeV^2$ for $A_1$ has been shifted to the right for the
sake of clarity.
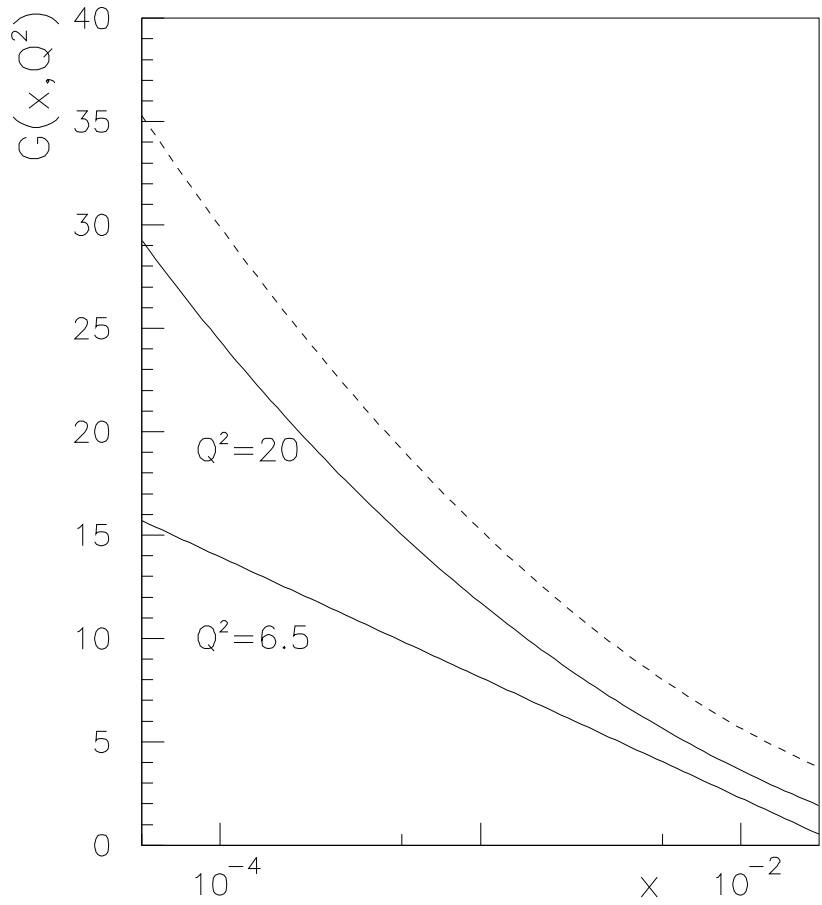
\item{fig. 2:} The gluon momentum density in the model, as function
of $x$, at $Q^2=6.5\,GeV^2$  and evolved at $Q^2=20\,GeV^2$
(continuous line). 
Dashed line shows the result of a leading order calculation~\cite{KP}.
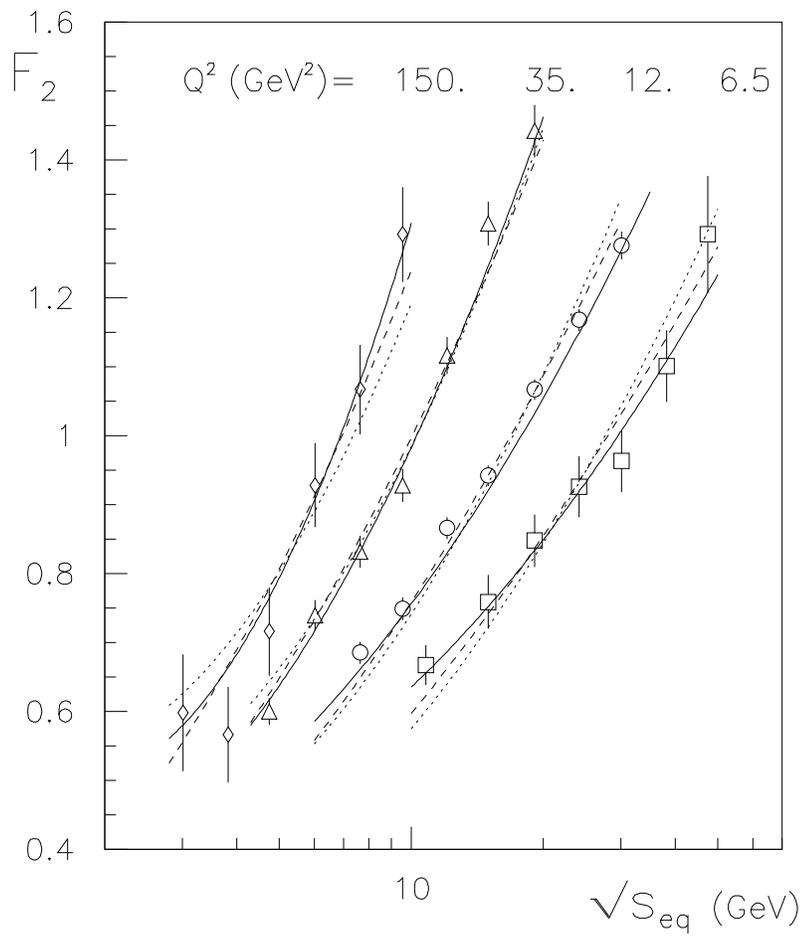
\item{fig. 3:} Proton structure function as function of
$\sqrt{s_{eq}}$ (see text) at different $Q^2$ values, 
data are from~\cite{H1} and continuous lines from our model.
Predictions obtained from the mean charged multiplicity,
normalized to $F_2(x,Q^2)$, are from~\cite{THO} (dashed line)
and from~\cite{ACT} (dotted line).

\end{description}

\end{document}